\documentstyle[a4,11pt]{article}

\newcommand{\nit}{\noindent}
\newcommand{\nl}{\newline}
\newcommand{\np}{\newpage}
\newcommand{\dsp}{\displaystyle}
\newcommand{\vs}[1]{\vspace{#1 ex}}
\newcommand{\hs}[1]{\hspace{#1 em}}
\newcommand{\bfr}{\begin{flushright}}
\newcommand{\efr}{\end{flushright}}
\newcommand{\bc}{\begin{center}}
\newcommand{\ec}{\end{center}}
\newcommand{\ben}{\begin{enumerate}}
\newcommand{\een}{\end{enumerate}}

\newcommand{\be}{\begin{equation}}
\newcommand{\ee}{\end{equation}}
\newcommand{\ba}{\begin{array}}
\newcommand{\ea}{\end{array}}
\newcommand{\ct}{\cite}
\newcommand{\bit}{\bibitem}
\newcommand{\dd}[2]{\frac{\partial{#1}}{\partial{#2}}}

\newcommand{\gam}{\gamma}
\newcommand{\del}{\delta}

\newcommand{\ve}{\varepsilon}

\newcommand{\thg}{\theta}
\newcommand{\kg}{\kappa}
\newcommand{\lb}{\lambda}

\newcommand{\vf}{\varphi}
\newcommand{\og}{\omega}
\newcommand{\Gam}{\Gamma}
\newcommand{\Del}{\Delta}

\newcommand{\lh}{\left(}
\newcommand{\rh}{\right)}
\newcommand{\ld}{\left.}
\newcommand{\rd}{\right.}

\newcommand{\der}{\partial}

\begin{document}

\pagestyle{empty}
\begin{flushright}
NIKHEF 01-012
\end{flushright} 

\begin{center}
{\Large{ \bf{Worldine deviations and epicycles} }}\\
\vs{5}

{\large J.W.\ van Holten} \\
\vs{2}

{\large{NIKHEF, Amsterdam NL}} \\
\vs{2}
{\tt t32@nikhef.nl}
\vs{2}

September 3, 2001\\
\vs{10}

{\small{ \bf{Abstract} }} \\
\end{center}

\nit
{\footnotesize{ In general relativity, only relative acceleration has 
an observer-independend meaning: curvature and non-gravitational 
forces determine the rate at which world lines of test bodies diverge 
or converge. We derive the equations governing both in the conventional 
geometric formalism as well as using the background field method.
This allows us to generalize the results to test bodies with 
charge and/or spin. The application of the equations to the motion 
of particles in a central field results in an elegant, fully 
relativistic version of the Ptolemaean epicycle scheme. 
}}

\np
\pagestyle{plain} 
\pagenumbering{arabic}

\nit
{\bf 1.\ World line deviation equations} 
\vs{1} 

\nit
According to the equivalence principle, structureless test bodies 
(sometimes referred to as point masses) in a gravitational field move 
on geodesics of space-time. Their worldline $x^{\mu}(\tau)$ is 
a solution of the geodesic equation 
\be 
\frac{D^2 x^{\mu}}{D\tau^2}\, =\, \frac{d^2 x^{\mu}}{d\tau^2}\, 
 + \Gam_{\lb\nu}^{\;\;\;\;\mu} \frac{dx^{\lb}}{d\tau}\,
 \frac{dx^{\nu}}{d\tau}\, =\, 0, 
\label{1}
\ee 
where the world-line parameter $\tau$ is to be taken as proper 
time. Introducing the four-velocity as the time-like tangent unit 
vector to the worldline: $u^{\mu} = dx^{\mu} / d\tau$, the equation 
can be written in geometrical language as 
\be 
u \cdot \nabla u = 0, \hs{2} u^2 = -1. 
\label{2}
\ee 
with $\nabla$ the covariant derivative. It is easily observed 
from eq.(\ref{1}), that the proper acceleration $a^{\mu} = d^2
x^{\mu} / d\tau^2$ is not a covariant object. In particular, its 
vanishing or non-vanishing has no observer-independent meaning. 

In contrast, the {\em relative} acceleration between worldlines is
a covariant quantity, and its vanishing or non-vanishing does not 
depend on the frame of reference \ct{synge}. We recall the argument.
Consider a one-parameter congruence of geodesics $x^{\mu}(\tau;\lb)$,
where $\lb$ labels the geodesics and $\tau$ is the proper-time 
parameter along the geodesic. We suppose the parametrization to 
be smooth, hence we can construct the tangent vector fields $u^{\mu} 
= \der x^{\mu}/\der \tau$, and $n^{\mu} = \der x^{\mu}/\der \lb$. 
It is straightfowardly established that 
\be 
(u \cdot \nabla n)^{\mu} = \frac{\der^2 x^{\mu}}{\der \tau \der \lb}\, 
 + \Gam_{\lb\nu}^{\;\;\;\;\mu} \dd{x^{\lb}}{\tau} \dd{x^{\nu}}{\lb} 
 = (n \cdot \nabla u)^{\mu}. 
\label{3}
\ee 
As a corrolary, we obtain
\be 
\ba{lll}
u \cdot \nabla(u \cdot \nabla n) & = & u \cdot \nabla (n \cdot \nabla u) 
 = (u \cdot \nabla n) \cdot \nabla u + u^{\mu} n^{\nu} 
 (\nabla_{\mu} \nabla_{\nu} u) \\
 & & \\
 & = & (n \cdot \nabla u) \cdot \nabla u + u^{\mu} n^{\nu} (\nabla_{\mu} 
 \nabla_{\nu} u) \\
 & & \\
 & = & n \cdot \nabla (u \cdot \nabla u) + 
 u^{\mu} n^{\nu} [\nabla_{\mu}, \nabla_{\nu}] u = u^{\mu} n^{\nu} 
 R_{\mu\nu}[u,\cdot]. 
\ea
\label{4} 
\ee
In component notation this reads
\be 
\frac{D^2n^{\mu}}{D\tau^2}\, =\, R_{\kg\lb\nu}^{\;\;\;\;\;\;\mu} 
 u^{\kg} u^{\nu} n^{\lb}.
\label{5}
\ee 
The interest in the deviation vector $n^{\mu}$ obviously derives 
from the fact that, if $x^{\mu}_0(\tau) = x^{\mu}(\tau;\lb_0)$ is 
a solution of the geodesic equation (\ref{1}), then to first order 
$x_1^{\mu} = x_0^{\mu} + n^{\mu} \Del \lb$ is a solution as well:
\be 
x^{\mu}(\tau;\lb_1) = x^{\mu}(\tau;\lb_0) + \Del \lb\, 
 \dd{x^{\mu}}{\lb}(\tau,\lb_0)\, \approx x^{\mu}(\tau;\lb_0 
 + \Del \lb). 
\label{6}
\ee 
It follows, that eq.(\ref{5}) describes the covariant relative 
acceleration between these world lines. Of course, $n^{\mu}$ is 
only a first approximation to the neighboring geodesic at $\lb_1 
= \lb_0 + \Del \lb$. To increase the precision of the approximation, 
one has to compute higher-order derivatives w.r.t.\ $\lb$, by 
solving higher-order versions of eq.(\ref{5}), involving not only 
the Riemann curvature tensor, but its derivatives as well. A 
systematic procedure of this type has been developed in ref.\ct{khc}.
Here I pursue the first-order equation (\ref{5}) and study some 
generalizations and applications. 

We first observe, that eq.(\ref{5}) is linear and homogeneous in
$n^{\mu}$. It is therefore not very difficult to construct an 
action from which it can be derived. The lagrangean of interest 
reads 
\be 
L(n) = \frac{1}{2}\, g_{\mu\nu} \frac{Dn^{\mu}}{D\tau} 
 \frac{Dn^{\nu}}{D\tau}\, + \frac{1}{2}\, R_{\mu\kg\nu\lb}\,
 u^{\kg} u^{\lb} n^{\mu} n^{\nu}. 
\label{7}
\ee 
In this lagrangean the metric, connection and curvature are those on 
the given reference geodesic $x_0^{\mu}(\tau)$, with $u^{\mu}(\tau) 
= \dot{x}^{\mu}_0$ representing the four-velocity along this same 
geodesic. These quantities act as background variables. Only the 
$n^{\mu}(\tau)$ are independent lagrangean generalized coordinates 
which are to be varied in the action. 

The action (\ref{7}) can be derived independently by starting 
from the geodesic lagrangean
\be 
L(x) = \frac{1}{2}\, g_{\mu\nu}(x)\, \frac{dx^{\mu}}{d\tau}
 \frac{dx^{\nu}}{d\tau}, 
\label{8}
\ee 
and expanding $x^{\mu}(\tau)$ near the given background geodesic
solution in the form $x^{\mu} = x_0^{\mu} + n^{\mu} \Del \lb$. 
The term independent of $\Del \lb$ does not contain $n^{\mu}$, 
and contributes a constant to the action. Next all terms 
linear in $\Del \lb$ drop out of the result because $x_0$ is a 
solution of the geodesic equation. Finally, the terms quadratic 
in $\Del \lb$ reproduce the expression (\ref{7}), up to a total
proper-time derivative and terms which vanish because of the 
geodesic equation for $x_0^{\mu}(\tau)$. Thus the lagrangean 
(\ref{7}) represents the lowest-order non-trivial term in a 
systematic expansion:
\be
S[x] = m \int d\tau L(x_0) + m (\Del \lb)^2 \int d\tau L(n) + 
 {\cal O}[(\Del \lb)^3]
\label{9}
\ee  
Obviously, the higher-order approximations can also be derived 
in this way. 

Clearly, as the variation of $L(n)$ w.r.t.\ $n^{\mu}$ reproduces 
the geodesic deviation equation (\ref{5}), the background-field 
method provides an alternative derivation of this equation. 
\vs{1} 

\nit
The above procedures can be generalized quite straightforwardly 
to cases in which test bodies are not completely structureless 
point masses, but carry e.g.\ charge and/or spin. In these 
cases particles do not move on geodesics, but on more general 
world lines \ct{jwvh,krip}. For the case of charged particles in 
a combined electro-magnetic and gravitational field, the resulting 
world line deviation equation was derived along the lines of 
eqs.(\ref{3})-(\ref{5}) in ref.\ct{bhk}. An alternative derivation 
using the background field method starts from the action 
\be 
S_q[x] = \int d\tau\, \lh \frac{m}{2}\, g_{\mu\nu}(x)\, 
 \dot{x}^{\mu} \dot{x}^{\nu} + q\, A_{\mu}(x)\, \dot{x}^{\mu} \rh,
\label{10}
\ee 
with the overdot the usual short-hand for proper-time derivatives. 
The world-lines given by the stationary points of this action are 
solutions of the Einstein-Lorentz equation 
\be 
\frac{D^2 x^{\mu}}{D\tau^2}\, = \frac{q}{m}\, F^{\mu}_{\;\;\nu}\,
 \frac{dx^{\nu}}{d\tau}. 
\label{11}
\ee 
Now given a solution $x_0^{\mu}(\tau)$ of this equation, and 
expanding the path in $S_q[x]$ as
\be 
x^{\mu}(\tau) = x_0^{\mu}(\tau) + \Del \lb\, n^{\mu}(\tau), 
\label{12}
\ee 
the action can be expanded to second order in $\Del \lb$ as 
\be 
\ba{lll}
S_q[x] & = & \dsp{ S_q[x_0] + \frac{(\Del \lb)^2}{2}\, 
 \int d\tau \left[ m \lh g_{\mu\nu}\, \frac{Dn^{\mu}}{D\tau} 
 \frac{Dn^{\nu}}{D\tau}\, + R_{\mu\kg\nu\lb} u^{\kg} u^{\lb} 
 n^{\mu} n^{\nu} \rh \rd }\\
 & & \\
 & & \dsp{\ld +\, q \lh F_{\mu\nu}\, n^{\mu} 
 \frac{Dn^{\nu}}{D\tau}\, + \nabla_{\mu} F_{\nu\lb}\, u^{\lb} 
 n^{\mu} n^{\nu} \rh \right]\, + {\cal O}[(\Del \lb)^3].}
\ea 
\label{13}
\ee 
To this order we then find that other solutions of the 
world-line equation (\ref{11}), close to $x_0^{\mu}(\tau)$, 
are given by (\ref{12}), with $n^{\mu}$ the solution of the 
world-line deviation equation \ct{bhk,kmmh}
\be 
\frac{D^2 n^{\mu}}{D\tau^2}\, =\, R_{\lb\nu\kg}^{\;\;\;\;\;\mu}
 u^{\lb} u^{\kg} n^{\nu} + \frac{q}{m}\, F^{\mu}_{\;\;\nu}\, 
 \frac{Dn^{\nu}}{D\tau}\, +\, \frac{q}{m}\, \nabla_{\lb} 
 F^{\mu}_{\;\;\nu}\, u^{\nu} n^{\lb}. 
\label{14}
\ee 
The alternative interpretation of $n^{\mu}$, as parametrizing the 
distance between two particles on neighboring world lines, holds 
in this case as well, provided the particles have the same 
charge-to-mass ratio $q/m$. Interestingly, this observation is not 
contradicted by the fact that one can obtain the Einstein-Lorentz
equation as well as the world-line deviation equation (\ref{14}) 
from reduction of the geodesic equation and geodesic deviation 
equation in five-dimensional space-time, as particles with different 
charge-to-mass ratio in four dimensions correspond to particles with 
different momentum in five-dimensional space-time \ct{kmmh}. 

Similarly, pseudo-classical spinning particles can be described by 
the supersymmetric lagrangean \ct{bcl,bdvh,jwvh}
\be 
L_{spin}(x,\psi) = \frac{1}{2}\, g_{\mu\nu} \dot{x}^{\mu} 
 \dot{x}^{\nu} + \frac{i}{2}\, \psi_a \frac{D\psi^a}{D\tau}, 
\label{15}
\ee 
with $\psi^a$ an anti-commuting tangent-space vector\footnote{The 
transition between base-space and tangent-space vectors is made 
as usual by the vierbein $e_{\mu}^{\;\;a}$ and its inverse.} such 
that the pseudo-classical spin is described by $S^{ab} = -i \psi^a 
\psi^b$. The corresponding equations of motion for spinning particles 
can be written as 
\be 
\frac{D^2 x^{\mu}}{D\tau^2}\, =\, \frac{1}{2}\, S^{ab} 
 R^{\mu}_{\;\;\nu ab} u^{\nu}, \hs{2} \frac{DS^{ab}}{D\tau} = 0.
\label{15.1}
\ee 
Starting from a one-parameter congruence of solutions 
$(x^{\mu}(\tau;\lb), \psi^a(\tau;\lb))$ we define the deviation 
vectors 
\be 
n^{\mu} = \dd{x^{\mu}}{\lb}, \hs{2} \xi^a = \frac{D\psi^a}{D\lb} 
 = \dd{\psi^a}{\lb} - n^{\mu}\, \og_{\mu\;\;b}^{\;\;a}\, \psi^b, 
\label{15.2}
\ee 
where $\og_{\mu\;\;b}^{\;\;a}$ is the spin connection. The 
covariant change in the spin-tensor is then
\be 
J^{ab} = \frac{DS^{ab}}{D\lb} = -i \lh \psi^a \xi^b + \xi^a \psi^b \rh.
\label{15.2.1}
\ee
These vectors satisfy the world-line deviation equations
\be 
\ba{lll}
\dsp{ \frac{D^2n^{\mu}}{D\tau^2} }& = & \dsp{ 
 R_{\kg\nu\lb}^{\;\;\;\;\;\mu} u^{\kg} u^{\lb} n^{\nu} +
 \frac{1}{2}\, S^{ab} R^{\mu}_{\;\;\nu ab} \frac{Dn^{\nu}}{D\tau} 
 }\\
 & & \\
 & & \dsp{ +\, \frac{1}{2} \lh S^{ab} \nabla_{\lb} 
 R^{\mu}_{\;\;\nu ab}\, u^{\nu} n^{\lb} + J^{ab} R^{\mu}_{\;\;\nu ab}\, 
 u^{\nu}\rh. }\\
 & & \\ 
\dsp{ \frac{DJ^{ab}}{D\tau} }& = & \dsp{ \left[ S, R_{\mu\nu} \right]^{ab} 
 u^{\mu} n^{\nu}. }
\ea  
\label{18}
\ee 
They define the stationary points of the quadratic deviation action 
\be 
\ba{lll}
L_{spin}(n,\xi) & = & \dsp{ \frac{1}{2}\, g_{\mu\nu} \frac{Dn^{\mu}}{D\tau} 
 \frac{Dn^{\nu}}{D\tau}\, + \frac{i}{2}\, \xi_a \frac{D\xi^a}{D\tau}  
 + \frac{1}{2}\, R_{\mu\kg\nu\lb} u^{\kg} u^{\lb} n^{\mu} n^{\nu} }\\
 & & \\ 
 & & \dsp{ -\, \frac{i}{4}\, \psi^a  \psi^b \lh R_{\mu\nu ab}\, n^{\mu} 
 \frac{Dn^{\nu}}{D\tau} + \nabla_{\mu} R_{\nu \lb ab}\, u^{\lb} n^{\mu}
 n^{\nu} \rh  }\\
 & & \\ 
 & & \dsp{ -\, i R_{\mu\nu ab}\, n^{\mu} u^{\nu} \xi^a \psi^b. }
\ea 
\label{19} 
\ee 

\nit
{\bf 2.\ Application: the Coulomb-Reissner-Nordstrom field} 
\vs{1}

\nit 
World-line deviation equations can be used to compute the relative 
motion between particles in given background fields, or to obtain 
an approximation to solutions for orbits close to a known one. 
We illustrate the general results with an application to the study 
of the motion of charged particles in a central gravitational and 
electric Coulomb-Reissner-Nordstrom field.

The vector potential and electric field strength for the Coulomb
part of this solution of the Einstein-Maxwell equations are
given by the one- and two-forms
\be 
A = - \frac{Q}{4\pi r}\, dt, \hs{2} 
F = dA = \frac{Q}{4\pi r^2}\, dr \wedge dt,
\label{2.1}
\ee 
whilst the metric for the gravitational field can be taken as 
\be 
- d\tau^2 = -B(r) dt^2 + \frac{1}{B(r)}\, dr^2 + r^2 \lh d\thg^2
 + \sin^2 \thg\, d\vf^2 \rh,
\label{2.2}
\ee 
where $B(r) = 1  - (2M/r) + (Q^2/r^2)$; $Q$ and $M$ are the charge 
and mass of the central body which is the source of the field. 

The orbits of particles with mass $m$ and charge $q$ in this 
background can be computed in closed form in terms of elliptic 
integrals \ct{bhk}. More precisely, the orbits are given by 
\be 
r(\vf) = \frac{r_0}{1 + e \cos y(\vf)}, 
\label{2.3}
\ee 
where $y(\vf)$ is the solution of the differential equation
\be 
\frac{dy}{d\vf}\, =\, \sqrt{A + B \cos y + C \cos^2 y},
\label{2.4}
\ee 
with coefficients of given by  
\be 
\ba{lll} 
A & = & \dsp{ 1 + \frac{Q^2}{\ell^2} \lh 1 - \lh \frac{q}{4\pi m} 
 \rh^2 \rh - \frac{6M}{r_0} + \frac{Q^2}{r_0^2} (6 + e^2), }\\
 & & \\
B & = & \dsp{ -\frac{2e}{r_0} \lh M - \frac{2Q^2}{r_0} \rh, }\\ 
 & & \\
C & = & \dsp{ \frac{e^2 Q^2}{r_0^2}. } 
\ea 
\label{2.5}
\ee 
Here $\ell$ is the constant angular momentum per unit of mass. As the 
periastra of the orbit are at the points $y(\vf) = 2\pi n$, one can now 
compute the angular distance $\Del \vf$ between successive periastra. 
Writing $\Del \vf = 2\pi + \del \vf$, it follows that the periastron 
shift per orbit is 
\be 
\del \vf = 2\pi \lh \frac{3M}{r_0} - \frac{Q^2}{2Mr_0} \rh + ...
\label{2.6}
\ee 
the dots denoting terms of higher order in $e$, $M/r_0$ or $Q/r_0$. 

Eqs.(\ref{2.3}) and (\ref{2.4}) describe a general orbit in the 
exterior region of the central body. However, they do not provide all 
information about the orbit. In particular, as the time coordinate 
has been eliminated from these equations, the solution does not 
tell us where in its orbit the test particle is at any moment. 
Such information can be relevant for some important applications, 
e.g.\ to compute estimates of the amount of electro-magnetic and 
gravitational radiation emitted by the system. The method of 
world-line deviations is useful to obtain parametrized expressions
of orbits $(r(t), \vf(t))$. 

As the reference orbit, the zeroth order approximation to the real orbit,
we take a circular one with constant radial coordinate $R$. Constants 
of motion on all orbits are the angular momentum per unit of mass, 
$\ell = \og R^2$, with $\og = \dot{\vf}$ the angular velocity, and 
the energy per unit of mass $\ve$, defined by 
\be
\frac{dt}{d\tau} = \frac{\ve - qQ/4\pi mR}{1 - 2M/R\, + Q^2/R^2}.
\label{2.7}
\ee 
On circular orbits the constants $R$, $\ell$ and $\ve$ are then 
related by  
\be  
\lh \ve - \frac{qQ}{4\pi mR} \rh^2 = \lh 1 - \frac{2M}{R} 
 + \frac{Q^2}{R^2} \rh\, \lh 1 + \frac{\ell^2}{R^2} \rh, 
\ee
and 
\be 
\ba{l}
\dsp{ \left[ \frac{\ell^2}{R} \hs{-.1} - M \hs{-.2} \lh 1 + 
 \frac{3\ell^2}{R^2} \rh \hs{-.2} + \hs{-.1} \frac{Q^2}{R} \hs{-.1} 
 \lh 1 + \frac{2\ell^2}{R^2} \rh \right]^2 \hs{-.5} = \hs{-.2} \lh 
 \frac{qQ}{4\pi m} \rh^2 \hs{-.3} \lh 1 + \frac{\ell^2}{R^2} \rh \hs{-.3} 
 \lh 1 - \frac{2M}{R} + \frac{Q^2}{R^2} \rh. }
\ea 
\label{2.8}
\ee 
As all orbits are planar, we can always choose the orientation of
the coordinate system such that $\thg = \pi/2$ for the reference orbit. 
For orbits tilted w.r.t.\ this one, we then find from eq.(\ref{14}) that 
\be 
\ddot{n}^{\thg} + \og^2 n^{\thg} = 0,
\label{2.9}
\ee 
from which it follows, as the physics dictates, that the distance 
perpendicular to the plane of the reference orbit oscillates with the 
period of the circular orbit $T = 2\pi/\og = 2\pi R^2/\ell$. 

Considering orbits in the plane of the reference orbit, the world-line
deviation equations (\ref{14}) for the other components $n^i = (n^t, 
n^r, n^{\vf})$ become
\be 
\ddot{n}^i + \gam^i_{\;j} \dot{n}^j + m^i_{\;j} n^j = 0,
\label{2.10}
\ee 
where the coefficient matrices take the form 
\be 
\gam = \lh \ba{ccc} 0 & \gam^t_{\;r} & 0 \\ 
                    \gam^r_{\;t} & 0 & \gam^r_{\;\vf} \\
                    0 & \gam^{\vf}_{\;r} & 0 \ea \rh, \hs{2} 
m = \lh \ba{ccc} 0 & 0 & 0 \\
                 0 & m^r_{\;r} & 0 \\
                 0 & 0 & 0 \ea \rh. 
\label{2.11}
\ee 
This represents a system of coupled linear oscillators, which 
has solutions 
\be 
n^t(\tau) = n^t_0 \sin \og_1 \tau, \hs{2} 
n^r(\tau) = n^r_0 \cos \og_1 \tau, \hs{2} 
n^{\vf}(\tau) = n^{\vf}_0 \sin \og_1 \tau,
\label{2.12}
\ee 
where $\og_1$ is the solution of the characteristic equation for 
(\ref{2.10}). The detailed form of this equation, using explicit 
expressions for the elements of the matrices $\gam$ and $m$ 
were given in \ct{bhk}. The resulting expression for the 
characteristic frequency is 
\be 
\og_1 = \og \lh 1 - \frac{3M}{R} + \frac{Q^2}{2MR} + ... \rh, 
\label{2.13}
\ee 
where the dots represent terms of higher order in $M/R$, $Q/M$ or
$q/m$. We also observe, that the amplitudes $n_0^i$ are not
all independent: as $u^2 = -1$ both on the original orbit and
on the displaced world-line, it follows that $n$ is space-like
and $u \cdot n = 0$. In the present case this general result 
translates to the constraint
\be 
\lh \ve - \frac{qQ}{4\pi mR} \rh n_0^t - \frac{qQ}{4\pi m \og_1 R^2}
 u^t n_0^r - \ell n_0^{\vf} = 0.
\label{2.14}
\ee 
Although the components $n^{\mu}$ define the direction of the deviation,
they do not determine the actual distance between neighboring world 
lines; this is given by equation (\ref{12}) as $\Del x^{\mu} = n^{\mu} 
\Del \lb$. Therefore, for any particular orbit specified by the circular
reference orbit (zeroth order approximation) and a world-line deviation 
vector $n$ (first order approximation), we must determine in addition 
the scale factor $\Del \lb$ to be applied. This can be done as follows. 
Comparing the approximate solution (\ref{2.10}) with the exact solution 
(\ref{2.3}), we observe that 
\be 
r(\vf) = R + \Del r \approx R - e R \cos y(\vf). 
\label{2.15}
\ee 
Hence at the periastron, one has 
\be 
\Del r = - \Del \lb n_0^r = - e R. 
\label{2.16}
\ee 
Thus the scale is set by the eccentricity of the orbit. 

%%Begin InstantTeX Picture 
\let\picnaturalsize=N 
\def\picsize{2.5in} 
\def\picfilename{epi.eps} 
%If you do not have the picture file add: 
%\let\nopictures=Y 
%to the beginning of the file. 
\ifx\nopictures Y\else{\ifx\epsfloaded Y\else\input epsf \fi 
\let\epsfloaded=Y 
\centerline{\ifx\picnaturalsize N\epsfxsize \picsize\fi \epsfbox{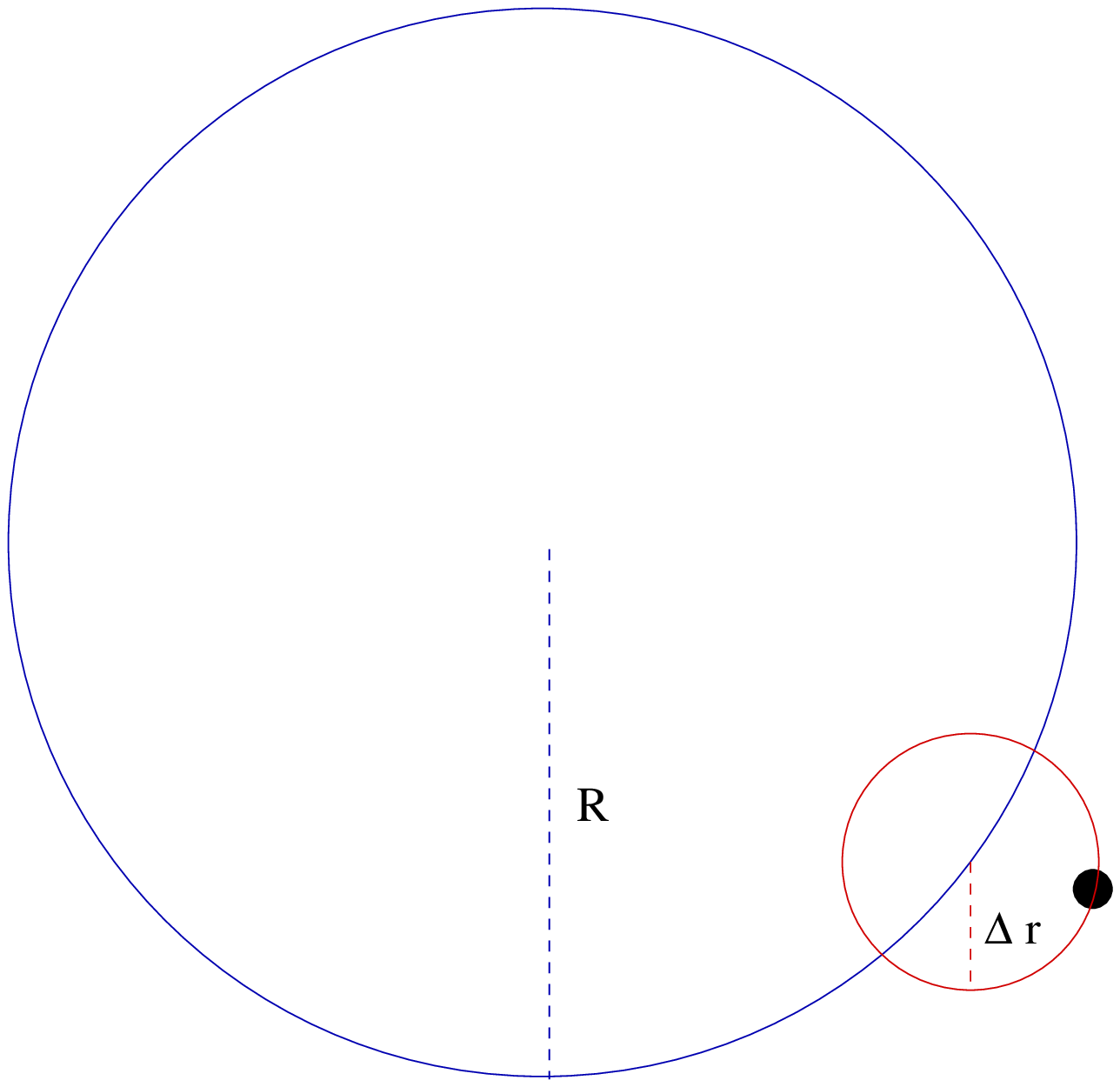}}}\fi 
%%End InstantTeX Picture 
\begin{center} 
{\footnotesize{ Fig.1: Epicycle approximation to orbit with eccentricity
$e = |\Del r|/R$.}}
\end{center} 

\nit
Finally, we can determine the shift in angular coordinate between 
successive periastra, i.e.\ the advance of the periastron per 
orbital period. First observe, that the periastron occurs at the 
minima of $n^r(\tau)$, i.e.\ for $\tau_n = (2n + 1) \pi /\og_1$. 
Thus the amount of proper time elapsing between periastra is 
$\Del \tau = 2 \pi / \og_1$; the corresponding period of observer
time is 
\be 
T = \int_0^{2\pi /\og_1} d\tau\, \frac{dt}{d\tau} 
 = \int_0^{2\pi /\og_1} d\tau\, (u^t + \dot{n}^t \Del \lb)
 = \frac{2\pi}{\og_1}\, u^t.
\label{2.17}
\ee 
Here $u^t$ is the rate of change of $t$ per unit of proper time 
along the circular reference orbit. Next we observe, that at 
the proper times $\tau_n$ the angular coordinates at the 
reference orbit and the true orbit coincide: $n^{\vf}(\tau_n) = 0$.
Hence the change in angular coordinate $\vf$ between successive 
periastra is the same as the change of this coordinate along the
circular reference orbit after time $T$. This we can easily 
compute. Defining
\be 
\del \vf = \vf(t_0 + T) - \vf(t_0) - 2 \pi,
\label{2.18}
\ee 
and using for the angular velocity the expression $d\vf/dt = 
\dot{\vf} d\tau/dt = \og/u^t$, we find 
\be 
\del \vf = \frac{\og T}{u^t} - 2 \pi = 2 \pi \lh \frac{\og}{\og_1} 
 - 1 \rh \approx 2 \pi \lh \frac{3M}{R} - \frac{Q^2}{2MR} \rh. 
\label{2.19}
\ee 
This is in perfect agreement with the expression (\ref{2.6}) 
obtained from the analytical form of the orbit. 

It is of interest to consider the geometrical interpretation of
the approximation scheme we have used in a little more detail. 
The zeroth order approximation to the orbit we have constructed
is a fully relativistic circular solution of the Einstein-Lorentz
equation in a Coulomb-Reissner-Nordstrom field, with period 
$T_0 = 2\pi /\og_0$. Included in this result is of course the 
simpler case of a circular geodesic in a Schwarzschild field. The 
first-order correction is a geodesic deviation which oscillates in 
all its components in the same plane with period $T_1 = 2 \pi/\og_1$.
Geometrically this represents another circular movement on the 
background of the zeroth-order solution, i.e.\ an {\em epicycle}, 
with period slightly different from the zeroth-order approximation. 
This has two immediate consequences: the orbit becomes 
eccentric, and the period between extrema of the orbit differs 
from the period of the average (zeroth order) circular motion.
This is in contrast with Newtonian gravity, where the periods 
are equal. Thus the extrema of the orbit (periastron and apastron) 
are shifted compared to the Newtonian approximation, by the amount 
predicted by the analytic description of the orbit. 

It can easily be shown \ct{khc}, that higher-order world-line 
deviations all satisfy linear harmonic-oscillator type 
equations. Thus, computing higher-order corrections to our
result amounts to the construction of higher-order epicycles. 
For the case of orbits in a central field, the method of 
world-line deviations then becomes a fully relativistic version 
of the Ptolemaean scheme \ct{pt}, which differs genuinely from 
the standard post-Newtonian approximation scheme because it uses 
the eccentricity of the orbit and the quantities $M/R$, $Q/M$ as 
expansion parameters, rather than $v/c$. As such this scheme 
offers an alternative to post-newtonian calculations of 
binary systems in a different physical regime, e.g.\ in the 
calculation of radiative effects. 
\vs{2}

\nit
{\bf Acknowledgement}\nl
I am indebted to Richard Kerner and Roberto Colistete for
enlightening ideas and enjoyable cooperation.

\end{document}